\documentclass[runningheads]{llncs}

\usepackage[T1]{fontenc}
\usepackage{amsmath,amsfonts}
\usepackage{graphicx}
\usepackage{algorithm}
\usepackage{algorithmic}
\usepackage{longtable}
\usepackage{color}
\usepackage[mathscr]{eucal}
\usepackage{xspace}
\usepackage{url}
\usepackage{tikz}
\usetikzlibrary{arrows,shapes,decorations,backgrounds,petri,mindmap,fit,positioning}
\usepackage{main}

\newcommand{\crit}[1]{\mathcal{#1}}
\newcommand{\critC}{\crit{C}}
\newcommand{\strat}{\sigma}
\newcommand{\fullstrat}{\pmb{\strat}}
\newcommand{\Strats}{\Sigma}
\newcommand{\In}{\mathit{In}}
\newcommand{\Best}{\mathit{Best}}
\newcommand{\Reach}{\mathit{Reach}}
\newcommand{\RDom}{\mathit{RDom}}

\newcommand{\Conflicts}{\mathit{Conflicts}}
\newcommand{\infosets}{\mathcal{I\!S}}

\begin{document}

\title{Pretty Good Strategies and Where to Find Them}
\author{Wojciech Jamroga\inst{1,2} \and
Damian Kurpiewski\inst{2,3}}
\institute{
    Interdisciplinary Centre for Security, Reliability and Trust, SnT, University of Luxembourg, Luxembourg \and
    Institute of Computer Science, Polish Academy of Sciences, Warsaw, Poland \and
Faculty of Mathematics and Computer Science, Nicolaus Copernicus University, Toru{\'n}, Poland
}

\maketitle

\begin{abstract}
Synthesis of bulletproof strategies in imperfect information scenarios is a notoriously hard problem.
In this paper, we suggest that it is sometimes a viable alternative to aim at ``reasonably good'' strategies instead.
This makes sense not only when an ideal strategy cannot be found due to the complexity of the problem, but also when no winning strategy exists at all.
We propose an algorithm for synthesis of such ``pretty good'' strategies.
The idea is to first generate a surely winning strategy with \emph{perfect information}, and then iteratively improve it with respect to two criteria of dominance: one based on the amount of conflicting decisions in the strategy, and the other related to the tightness of its outcome set.
We focus on reachability goals and evaluate the algorithm experimentally with very promising results.
\end{abstract}


\section{Introduction}\label{sec:intro}
As the systems around us become more complex, and at the same time more autonomous, the need for {unambiguous specification} and {automated verification} rapidly increases.
Many relevant properties of multi-agent systems refer to \emph{strategic abilities} of agents and their groups.
For example, functionality requirements can be often understood in terms of the user's ability to complete the selected tasks.
Similarly, many security properties boil down to inability of the intruder to obtain his goals.
\emph{Logics of strategic reasoning} provide powerful tools to reason about such aspects of MAS~\cite{Alur02ATL,Schobbens04ATL,Mogavero10stratLogic,Berthon17sl-imperfectinfo,BerthonMMRV21,Jamroga15specificationMAS}.
A typical property that can be expressed says that \emph{the group of agents $A$ has a collective strategy to enforce temporal property $\varphi$, no matter what the other agents in the system do}. In other words, $A$ have a ``winning strategy'' that achieves $\varphi$ on all its possible execution paths.

Specifications in agent logics can be then used as input to \emph{model checking}~\cite{Clarke81ctl}, which makes it possible to verify the correct behavior of a multi-agent system by an automatic tool~\cite{Alur00mocha,Dembinski03verics,Gammie04mck,Cermak14mcheckSL,Cermak15mcmas-sl-one-goal,Lomuscio17mcmas,Kurpiewski19stv-demo,Kurpiewski21stv-demo}.
Moreover, model checking of strategic formulas typically relies on synthesis of a suitable strategy to demonstrate that such a strategy exists.

Verification and reasoning about strategic abilities is difficult for a number of reasons. The
prohibitive complexity of model checking and strategy synthesis is a well
known factor~\cite{Bulling10verification,Doyen11games,Mogavero14behavioral,Berthon17sl-imperfectinfo,BerthonMMRV21}.
This can be overcome to some degree by using efficient symbolic methods and data
structures~\cite{Bryant86boolean,Burch90symbolic-mcheck,Gammie04mck,Raimondi07obdds}.
However, real-life agents typically have limited capabilities of observation, action, and reasoning.
That brings additional challenges. First, the theoretical complexity of
model checking for imperfect information strategies (sometimes called \emph{uniform strategies}) ranges from
$\NP$--complete to undecidable~\cite{Schobbens04ATL,Bulling10verification,Dima11undecidable}, depending on the precise setup of the problem.
Secondly, practical attempts at verification suffer from state-space and transition-space explosion.
Thirdly, there is no simple fixed-point characterisation of typical properties~\cite{Dima15fallmu,Bulling11mu-ijcai}.
As a consequence of the latter, most approaches to synthesis and verification
boil down, in the worst case, to checking all the possible strategies~\cite{Lomuscio06uniform,Busard15reasoning,Calta10finding,Pilecki17smc,Kurpiewski19domination}.
Unfortunately, the strategy space is huge -- usually larger than the state space by orders of magnitude, which makes brute-force search hopeless.

An interesting attempt at heuristic search through the strategy space has been proposed in~\cite{Kurpiewski19domination}.
There, a concept of domination between strategies was introduced, based on the ``tightness'' of the outcome sets induced by the strategies.
Formally, strategy $s$ dominates $s'$ if the set of possible executions of $s$ is a strict subset of the executions of $s'$.
The intuition is that those strategies are better which give the agent a better grip on what is going to happen, and better reduce the nondeterminism of the system.
Then, the authors of~\cite{Kurpiewski19domination} proposed an algorithm for synthesis of uniform strategies, based on depth-first search through the strategy space with simultaneous optimization of dominated partial strategies.
The algorithm, dubbed DominoDFS, performed with considerable success on several benchmarks.
This might have had two related reasons.
First, restricting the set of successor states reduces the possibility of encountering a ``bad'' successor further on.
Perhaps even more importantly, it reduces the space of reachable states, and hence has the potential to considerably speed up the computation.

In this paper, we take the idea of dominance-based optimization, and apply it from a completely different angle.
Most importantly, we propose that searching for a ``reasonably good'' strategy is sometimes a viable alternative to the search for an ideal one (where ``ideal'' means a surely winning imperfect information strategy).
This obviously makes sense when no winning strategy exists, but also when an ideal strategy cannot be found due to the complexity of the problem.
Moreover, we propose a procedure for synthesis of such ``pretty good'' strategies.
The algorithm starts with generating a surely winning strategy with \emph{perfect information}.
Then, it iteratively improves it with respect to two criteria of dominance: one based on the amount of conflicting decisions in the strategy, and the other related to the tightness of its outcome set.
It is worth noting that this is an \emph{anytime} algorithm. Thus, it \emph{always} returns some strategy, provided that a perfect information strategy has been generated in the first phase.

We evaluate the algorithm experimentally on randomly generated concurrent game structures with imperfect information, as well as the scalable Drones benchmark of~\cite{Kurpiewski19domination}.
The results are compared to the output of DominoDFS and to the fixpoint approximation algorithm of~\cite{Jamroga19fixpApprox-aij}, forming a very promising pattern.
In particular, for models with relatively small information sets (a.k.a. epistemic indistinguishability classes), our algorithm was able to find \emph{ideal} strategies where the other approaches consistently failed.
We note that, according to the theoretical results proposed in~\cite{Jamroga19somethings}, approaches relying on search through the space of uniform strategies may be feasible for models with {large} information sets. At the same time, they are unlikely to succeed for models with small epistemic classes.
This makes our new method a potentially good complement to algorithms like DominoDFS.

\paragraph{Outline of the Paper}
The structure of the paper is as follows.
We begin by introducing the standard semantics of strategic ability in Section~\ref{sec:prelim}.
We also cite the complexity results for model checking and strategy synthesis, and recall the notion of strategic dominance from~\cite{Kurpiewski19domination} that will serve as inspiration for our heuristics.
In Section~\ref{sec:dominance-two}, we propose an abstract template for multicriterial strategic dominance, and instantiate it by two actual dominance relations that will provide the heuristics.
Our algorithm for strategy synthesis based on iterated improvement is presented in Section~\ref{sec:algorithm}, and evaluated experimentally in Section~\ref{sec:experiments}.
We also discuss how the algorithm can be extended to synthesis of coalitional strategies in Section~\ref{sec:coalitions}.
Finally, we conclude in Section~\ref{sec:conclusions}.

\paragraph{Related Work}
A number of frameworks has been aimed at the verification of strategic properties under imperfect
information. Regarding the available tools, the state-of-the-art MAS model checker {MCMAS}~\cite{Lomuscio06uniform,Lomuscio17mcmas} combines
efficient symbolic representation of state-space using Binary Decision Diagrams (BDDs) with
exhaustive iteration over uniform strategies.
A similar approach based on exhaustive search through strategy space is presented in~\cite{Calta10finding}.
A prototype tool {SMC}~\cite{Pilecki17smc} employs bounded unfoldings of transition relation with strategy exploration and calls to {MCMAS}.
Strategy search with optimisation of partial strategies has been further used in~\cite{Busard15reasoning,Busard17phd,Kurpiewski19domination}.
Most relevant to us, the optimisation in~\cite{Kurpiewski19domination} was driven by strategic dominance based on the tightness of the outcome set.

Other recent attempts at feasible verification of uniform strategies include~\cite{Belardinelli17abstraction,Jamroga17bisimATLir,Jamroga18por} that propose methods for reduction of models with incomplete information, based respectively on abstraction, bisimulation, and partial-order equivalences.
Another method~\cite{Jamroga19fixpApprox-aij} avoids the brute-force strategy search by using fixpoint approximations of the input formulas.
A prototype tool STV implementing the DominoDFS algorithm and the fixpoint approximation was reported in~\cite{Kurpiewski19stv-demo}.

We note that all the above approaches try to directly synthesize an ideal (i.e., uniform surely winning) strategy for the given goal.
In contrast, our new algorithm starts with a flawed strategy (namely, surely winning but not uniform), and attempts to do iterative improvement.
As we show, this may well end up in producing an ideal solution in cases where the other methods are inconclusive.
No less importantly, our algorithm produces reasonably good strategies even when an ideal one cannot be found.
The only related work in model checking of multi-agent systems, that we are aware of, is~\cite{Bulling09patl-fundamenta,Aminof18gradedSL} where a theoretical framework was proposed for reasoning about strategies that succeed on ``sufficiently many'' outcome paths.

\section{Preliminaries}\label{sec:prelim}
In this section we recall the standard formal framework used for reasoning about strategies in MAS.
To this end, \emph{alternating-time temporal logic \ATL}~\cite{Alur97ATL,Alur02ATL,Schobbens04ATL} is often used.
We also recall the notion of dominance for partial strategies, that was proposed in~\cite{Kurpiewski19domination}.

\subsection{\ATL: What Agents Can Achieve}

\ATL~\cite{Alur97ATL,Alur02ATL,Schobbens04ATL} generalizes the branching-time temporal logic \CTL~\cite{Clarke81ctl} by replacing the path quantifiers $\Epath,\Apath$ with
\emph{strategic modalities} $\coop{A}$.
Formulas of \ATL allow to express intuitive statements about what agents (or groups of agents) can achieve.
For example, $\coop{W,E}\Sometm\,\prop{win_{WE}}$ says that the players West and Eeast in a game of Bridge can jointly win the game.
Formally, the syntax of \ATL is defined by the following grammar:
\[
\phi ::= p \mid \neg\phi \mid \phi\land\phi \mid \coop{A}\Next\phi
\mid \coop{A}\Always\phi \mid \coop{A}\phi \Until\phi,
\]
where $p \in \Props$ is an atomic proposition and $A\subseteq\Agt$ is a group of agents.
We read $\coop{A}\gamma$ as \emph{``$A$ can identify and execute a strategy that enforces $\gamma$,''}
$\Next$ as \emph{``in the next state,''} $\Always$ as \emph{``now and always in the future,''} and $\Until$ as \emph{``until.''}

\subsection{Models}

We interpret \ATL~\cite{Alur02ATL,Schobbens04ATL} specifications over a variant of transition systems where transitions are labeled with combinations of actions, one per agent. Moreover, epistemic relations are used to indicate states that look the same to a given agent.
Formally, an \emph{imperfect information concurrent game structure}, or simply a \emph{model}, is given by
$\model = \tuple{\Agt, \States, \Props, \V, Act, d, o,\{\sim_a \mid a\in \Agt\}}$
which includes a nonempty finite set of agents $\Agt = \set{1,\dots,k}$, a nonempty set of states $\States$, a set of atomic propositions $\Props$ and their valuation $\V\colon\Props\rightarrow \powerset{\States}$, and a nonempty finite set of (atomic) actions $\Actions$. The protocol function $\prot \colon \Agt \times \States \rightarrow \powerset{Act}$ defines nonempty sets of actions available to agents at each state; we will write $\prot_a(q)$ instead of $\prot(a,q)$, and define $\prot_A(q) = \prod_{a\in A}\prot_a(q)$ for each $A\subseteq\Agt, q\in\States$.
Furthermore, $o$ is a (deterministic) transition function that assigns the outcome state $q' = o(q,\alpha_1,\dots,\alpha_k)$ to each state $q$ and tuple of actions $\langle\alpha_1, \dots, \alpha_k\rangle$ such that $\alpha_i \in \prot_i(q)$ for $i=1, \dots, k$.
Every $\sim_a\subseteq \States\times\States$ is an epistemic equivalence relation with the intended meaning that, whenever $q\sim_a q'$, the states $q$ and $q'$ are indistinguishable to agent $a$.
By $[q]_a$ we mean the set of states indistinguishable to agent $a$ from the state $q$.
The model is assumed to be \emph{uniform}, in the sense that $q\sim_a q'$ implies $d_a(q)=d_a(q')$.
Note that perfect information can be modeled by assuming each $\sim_a$ to be the identity relation.

\subsection{Strategies}

A strategy of an agent $a\in\Agt$ is a conditional plan that specifies what $a$
is going to do in every possible situation. The details of the definition depend on
the observational capabilities of the agent and its memory. In this paper we consider
the case of \emph{imperfect information imperfect recall} strategies (sometimes also called \emph{uniform memoryless strategies}), where
an agent can observe only a part of the environment (i.e., perceives some states as
indistinguishable) and performs the same action every time a given state is reached.

Formally, a uniform strategy for $a$ is a function $\fullstrat_a \colon \States\to\Actions$ satisfying
$\fullstrat_a(q)\in\prot_a(q)$ for each $q\in\States$ and $\fullstrat_a(q) = \fullstrat_a(q')$
for each $q,q'\in\States$ such that $q\sim_a q'$.
A \emph{collective uniform strategy} $\fullstrat_A$ for a coalition $A\subseteq\Agt$
is a tuple of individual strategies, one per agent from $A$.

\subsection{Outcome Paths}

A \emph{path} $\lambda=q_0q_1q_2\dots$ is an infinite sequence of states such that there is a transition between each $q_i,q_{i+1}$.
We use $\lambda[i]$ to denote the $i$th position on path $\lambda$
(starting from $i=0$) and $\lambda[i,j]$ to denote the part of $\lambda$ between positions $i$ and $j$.
Function $out(q,\fullstrat_a)$ returns the set of all paths that can result from the execution of a strategy $\fullstrat_a$, beginning at state $q$. Formally:
\begin{description}
\item[$out(q,\fullstrat_a) =$]  $\{ \lambda=q_0,q_1,q_2\ldots \mid
      q_0=q$ and for each $i=0,1,\ldots$ there exists
      $\tuple{\alpha^{i}_{a_1},\ldots,\alpha^{i}_{a_k}}$ such that
      $\alpha^{i}_{a} \in d_a(q_{i})$ for every $a\in \Agt$,
      and $\alpha^{i}_{a} = \fullstrat_A|_a(q_{i})$ for every $a\in A$,
      and $q_{i+1} = o(q_{i},\alpha^{i}_{a_1},\ldots,\alpha^{i}_{a_k}) \}$.
\end{description}
\begin{sloppypar}
\noindent
Moreover, the function $out^{\ir}(q,\fullstrat_a) = \bigcup_{a\in A}\bigcup_{q\sim_a q'}\out(q',\fullstrat_a)$ collects all the outcome paths that start from states that are indistinguishable from $q$ to at least one agent in $A$.
\end{sloppypar}

\subsection{Semantics of \ATL}

Given a model $\model$ and a state $q$, the semantics of \ATL formulas is defined as follows:
\begin{itemize2}
\item $\model,q\satisf p$ iff $q\in\V(p)$,
\item $\model,q\satisf\neg\phi$ iff $\model,q\not\satisf\phi$,
\item $\model,q\satisf\phi\land\psi$ iff $\model,q\satisf\phi$ and $\model,q\satisf\psi$,
\item $\model,q\satisf\coop{A}\Next\phi$ iff there exists a uniform strategy $\fullstrat_A$ such that for all $\lambda\in\out^{\ir}(q,\fullstrat_A)$ we have $\model,\lambda[1]\satisf\phi$,
\item $\model,q\satisf\coop{A}\Always\phi$ iff there exists a uniform $\fullstrat_A$ such that for all $\lambda\in\out^{\ir}(q,\fullstrat_A)$
    and $i\in\Nat$ we have $\model,\lambda[i]\satisf\phi$,
\item $\model,q\satisf\coop{A}\psi\Until\phi$ iff there exists a uniform $\fullstrat_A$ such that for all $\lambda\in\out^{\ir}(q,\fullstrat_A)$ there is $i\in\Nat$ for which $\model,\lambda[i]\satisf\phi$ and $\model,\lambda[j]\satisf\psi$ for all $0\le j < i$.
\end{itemize2}
The standard boolean operators (logical constants $\top$ and $\bot$, disjunction $\lor$, and implication $\then$) are defined as usual.
Additionally, we define \emph{``now or sometime in the future''} as $\Sometm\varphi \equiv \top\Until\varphi$.
It is easy to see that $\model,q\satisf\coop{A} \Sometm\phi$ iff there exists a collective uniform strategy $\fullstrat_A$ such that, on each path $\lambda\in\out^{\ir}(q, \fullstrat_A)$, there is a state that satisfies $\phi$.

\subsection{Model Checking and Strategy Synthesis}

It is well known that model checking of \ATL based on uniform memoryless strategies is \Deltwo-complete with respect to the size of the explicit (global) model~\cite{Schobbens04ATL,Jamroga06atlir-eumas,Bulling10verification}, i.e., on top of the usual state-space and transition-space explosion which arises from the composition of the agents' local models.
This concurs with the results for solving imperfect information games and synthesis of winning strategies, which are also known to be hard~\cite{Chatterjee07omegareg,Doyen11games,Peterson79alternation}.
Note that model checking \ATL corresponds very closely to strategy synthesis for reachability/safety games.
In fact, most model checking algorithms for \ATL try to build a winning strategy when checking if such a strategy exists.

It is also known that both strategy synthesis and \ATL model checking for imperfect information are not only theoretically hard, they are also difficult in practice.
In particular, imperfect information strategies do not admit straightforward fixpoint algorithms based on standard short-term ability operators~\cite{Bulling11mu-ijcai,Dima14mucalc}. That makes incremental synthesis of strategies impossible, or at least difficult to achieve.
Some practical attempts to overcome the barrier have been reported in~\cite{Busard17phd,Busard14improving,Busard15reasoning,Huang14symbolic-epist,Pilecki14synthesis,Jamroga19fixpApprox-aij,Kurpiewski19domination}.
Up until now, experimental results confirm that the initial intuition was right: model checking of strategic modalities for imperfect information is hard, and dealing with it requires innovative algorithms and verification techniques.

We emphasize that, at the same time, model checking for perfect information strategies (i.e., ones that can specify different choices at indistinguishable states) is much cheaper computationally, namely \Ptime-complete in the size of the model~\cite{Alur02ATL}.

\subsection{Partial Strategies and Strategy Dominance}

A \emph{partial strategy} for $a$ is a partial function
$\strat_a \colon \States\partto\Actions$ that can be extended to a strategy.
The domain of a partial strategy is denoted by $\dom{\strat_a}$.
The set of all partial strategies for $A\subseteq\Agt$ is denoted by $\Strat_A$.

Let $q\in\dom{\strat_A}$ for some $\strat_A\in\Strat_A$.
The \emph{outcome} of $\strat_A$ from $q$ consists of all
the maximal paths $\lambda\in\dom{\strat_A}^{*}\cup\dom{\strat_A}^{\omega}$
that follow the partial strategy.
Formally we have:
\begin{align*}
\lambda\in\out(q,\strat_A) \text{ iff }
\lambda_1 = q
\land
\forall_{i\le\psize{\lambda}} \lambda_i\!\in\!\dom{\strat_A}\\
\land\;
\forall_{i<\psize{\lambda}}
\exists_{\actrest\in\prot_{\Agt\setminus A}(\lambda_i)}
\ofun(\lambda_i, (\strat_A(\lambda_i), \actrest)) = \lambda_{i+1}
\end{align*}
where $\psize{\lambda}$ denotes the length (i.e., the number of states) of
$\lambda$ and $\lambda$ is either infinite or cannot be extended.
For each $i\in\Nat$ let $\lambda_i$ denote the $i$--th element.
Let $Q\subseteq\dom{\strat_A}$.
A partial strategy $\strat_A$ is $Q$-\emph{loopless}, if the set
$\bigcup_{q\in Q}\out(q,\strat_A)$ contains only finite paths.
For each $\lpropositionp\in\Props$ we say that $\strat_A$ is
$\lpropositionp$-free if
$\V(\lpropositionp) \cap \dom{\strat_A} = \emptyset$.

In what follows, we often refer to partial strategies simply as strategies and assume a fixed \ICGS\
and $A\subseteq\Agt$.

The paper~\cite{Kurpiewski19domination} proposed a notion of strategic dominance defined with respect to a given context.
Assume that we want to compare two partial strategies $\strat_A$ and $\strat_A'$.
First, we fix a context strategy $\strat_A^C$, such that after executing it the control can be given to strategy $\strat_A$ or $\strat_A'$.
Then, we say that $\strat_A$ dominates $\strat_A'$, iff the sets of input states\footnote{i.e., initial states of the strategy} of both strategies are equal, and the set of output states of strategy $\strat_A$ is a subset of the set of output states of strategy $\strat_A'$.

\section{Two Notions of Dominance for Iterated Strategy Improvement}\label{sec:dominance-two}

In~\cite{Kurpiewski19domination}, partial strategies are optimized according to only one criterion, namely the tightness of their outcome sets.
In contrast, we propose to use two dimensions for optimization: tightness of the outcome \emph{and} uniformity of the actions selected within the strategy.
This is because, unlike~\cite{Kurpiewski19domination}, we start the synthesis with a perfect information strategy.
Thus, our algorithm optimizes strategies that can include any number of conflicts, in the sense that it might prescribe different actions within the same information set $[q]_a$.

\subsection{Multi-Criterial Domination: Abstract Template}

Consider a set of partial strategies $\Strats$ of agent $a$, based on the same epistemic class of $a$.
That is, there exists $q\in\States$ such that $dom(\strat)\subseteq [q]_a$ for every $\strat\in\Strats$.
Let $\strat_1,\strat_2 \in \Strats$.
We begin with an abstract definition of domination that looks at two criteria $\critC_1, \critC_2$. The idea is that $\strat_2$ dominates $\strat_1$ if it improves on $\critC_1$ without deteriorating with respect to $\critC_2$.

\begin{definition}[$(\critC_1, \critC_2)$-domination]
Let each $\critC_i$ be a criterion associated with a total order $\preceq_{\critC_i}$ on the partial strategies in $\Strats$.
The strict variant $\prec_{\critC_i}$ of the ordering is defined in the obvious way, by \mbox{$\preceq_{\critC_i} \setminus (\preceq_{\critC_i})^{-1}$}.
We say that \emph{$\strat_1$ is $(\critC_1, \critC_2)$-dominated by $\strat_2$} iff it holds that $\strat_1 \prec_{\critC_1} \strat_2$ and at the same time $\strat_1 \preceq_{\critC_2} \strat_2$.
\qed
\end{definition}

\begin{definition}[Better and best domination]
Consider partial strategies $\strat_2,\strat_2'$ that both $(\critC_1, \critC_2)$-dominate $\strat_1$.
We say that $\strat_2$ \emph{better $(\critC_1, \critC_2)$-dominates} $\strat_1$ iff $\strat_2' \prec_{\critC_1} \strat_2$, i.e., $\strat_2$ performs better than $\strat_2'$ with respect to the primary criterion $\critC_1$.
Note: the fact that $\strat_2$ better dominates $\strat_1$ than $\sigma_2'$ does not imply that $\strat_2$ dominates $\strat_2'$, because $\strat_2$ may perform poorer than $\strat_2'$ on the secondary criterion $\critC_2$.

Moreover, $\strat_2$ \emph{best dominates} $\strat_1$ with respect to $(\critC_1, \critC_2)$ iff it dominates $\strat_1$ and no other strategy in $\Strats$ better dominates $\strat_1$.
The set of strategies that best dominate $\strat_1$ with respect to $(\critC_1, \critC_2)$ will be denoted by $\Best_{\critC_1, \critC_2}(\sigma_1)$.
\qed
\end{definition}

\subsection{Outcome- and Uniformity-Dominance}

In the following, we assume a shared set of input nodes $In\subseteq dom(\strat_1), dom(\strat_2)$.
The set of states reachable from $\In$ by partial strategy $\sigma_i$ is denoted by $\Reach(\In,\strat_i)$.
Furthermore, we define the \emph{domain of relevance} of $\strat_i$ as $\RDom(\In,\strat_i) = dom(\strat_i) \cap \Reach(\In,\strat_i)$.
That is, $\RDom(\In,\strat_i)$ excludes from the domain of $\strat_i$ the states that cannot be reached, and hence are irrelevant when reasoning about potential conflicts between choices.

The \emph{outcome criterion} is given by relation $\preceq_{\crit{O}(\In)}$ such that
$\strat_1 \preceq_{\crit{O}(\In)} \strat_2$ iff $\Reach(\In,\strat_2) \subseteq \Reach(\In,\strat_1)$, i.e., $\strat_2$ has at least as tight set of reachable outcome states as $\strat_1$.

We will now proceed to the other criterion, related to uniformity of strategies.
First, we define the \emph{conflict set} of $\strat_i$ on states $Q\subseteq\States$ as $\Conflicts(Q,\strat_i) = \set{(q,q') \in Q\times Q \mid \strat_i(q)\neq\strat_i(q')}$, i.e., the set of all pairs of states from $Q$ where $\strat_i$ specifies conflicting choices. \\
Now, the \emph{uniformity criterion} is given by relation $\preceq_{\crit{U}(\In)}$ such that
$\strat_1 \preceq_{\crit{U}(\In)} \strat_2$ iff $\Conflicts(\RDom(\In,\strat_2),\ \strat_2) \subseteq \Conflicts(\RDom(\In,\strat_1),\ \strat_1)$.
In other words, all the conflicts that $\strat_2$ encounters in its domain of relevance must also appear in $\strat_1$ (but not necessarily vice versa).

\begin{definition}[Outcome- and uniformity-domination]
We say that \emph{$\strat_1$ is outcome-dominated by $\strat_2$ on input $\In$} iff it is $(\crit{O}(\In),\crit{U}(\In))$-dominated by $\strat_2$.
Likewise, \emph{$\strat_1$ is uniform-dominated by $\strat_2$ on input $\In$} iff it is $(\crit{U}(\In),\crit{O}(\In))$-dominated by $\strat_2$.
The concepts of better and best domination apply in a natural way.
\qed
\end{definition}

\section{Iterated Strategy Synthesis}\label{sec:algorithm}
\begin{algorithm}[t]
    \caption{Synthesis algorithm $strat\_synth(M)$}
    \begin{algorithmic}
    \STATE{Generate a winning perfect information strategy $\fullstrat$}
    \IF{$\fullstrat$ doesn't exist}
        \STATE{return \textbf{false}}
    \ENDIF
    \STATE{Create an empty list $PStr$}
    \STATE{Create a list $\infosets$ of information sets in $\update{M}{\fullstrat}$}
    \FOR{$i=1$ \TO $|\infosets|$}
        \STATE{Take the info set $(i,Q_{i})$ and generate the corresponding partial strategy $\strat_{i}$ as a restriction of $\fullstrat$ to $Q_{i}$ and add it to $PStr$}
        \STATE{$\In_{i} := Q_i\ \cap\ \Reach\big(\ \Reach(Q_0,\fullstrat) \setminus Q_{i},\ (\fullstrat\setminus\strat_{i})\ \big)$}
        \STATE{$\RDom_{i} := Q_i\ \cap\ \Reach(\In_i,\strat_i)$}
        \STATE{$Out_{i} := \Reach(\In_i,\strat_i)\ \setminus\ Q_i$}
        \STATE{$\Conflicts_{i} := \Conflicts(\RDom_i,\strat_i)$}
    \ENDFOR
    \STATE{Optimize the resulting list of partial strategies $PStr$}
    \RETURN{$PStr$}
    \end{algorithmic}
    \label{fig:alg-syn}
\end{algorithm}

\begin{algorithm}[t]
    \caption{Single sweep optimization algorithm $optimize\_once(PStr)$}
    \begin{algorithmic}
        \STATE{$OldPStr := PStr$}
        \FOR{$i=1$ \TO $|\infosets|$}
            \REPEAT
                \STATE{ $OldPStr_i := PStr(i)$}
                \IF{exists $\strat$ that uniform-best dominates $PStr(i)$ in $\In_i$}
                    \STATE{update $PStr(i)$ by taking $\strat_i := \strat$ and recomputing the sets $\RDom_{i}$, $Out_{i}$, and $\Conflicts_{i}$}
                \ENDIF
                \IF{exists $\strat$ that outcome-best dominates $PStr(i)$ in $\In_i$}
                    \STATE{update $PStr(i)$ by taking $\strat_i := \strat$ and recomputing the sets $\RDom_{i}$, $Out_{i}$, and $\Conflicts_{i}$}
                \ENDIF
            \UNTIL{$PStr(i) = OldPStr_i$}
            \STATE{update $\fullstrat$ with the current contents of $PStr$}
            \FOR{every $j\neq i$}
                \STATE{update the input states of $PStr(j)$ by $\In_{j} := Q_j\ \cap\ \Reach\big(\ \Reach(Q_0,\fullstrat) \setminus Q_{j},\ (\fullstrat\setminus\strat_{j})\ \big)$}
            \ENDFOR
        \ENDFOR
        \RETURN{$PStr$}
    \end{algorithmic}
    \label{fig:alg-once}
\end{algorithm}

\begin{algorithm}[t]
    \caption{Optimization algorithm $optimize(PStr)$}
    \begin{algorithmic}
        \REPEAT
            \STATE{$OldPStr := PStr$}
            \STATE{$Pstr := optimize\_once(PStr)$}
        \UNTIL{timeout or (PStr = OldPStr)}
        \RETURN{$PStr$}
    \end{algorithmic}
    \label{fig:alg}
\end{algorithm}

In this section, we propose an algorithm for strategy synthesis, based on the following idea: first generate a surely winning perfect information strategy (if it exists), and then iteratively improve it with respect to the dominance relations proposed in Section~\ref{sec:dominance-two}. Of the two relations, uniformity-dominance has higher priority.
The iterative improvement terminates when the procedure reaches a fixpoint (i.e., no more improvement is possible anymore) or when the time limit is exceeded.
After that, the optimized strategy is returned and checked for uniformity.

We will now define our procedure in more detail.

\begin{definition}[Input]
The input of the algorithm consists of: model $M$, state $q$ in $M$, and formula $\coop{a}\Sometm\varphi$.
We define the set of \emph{initial states} as $Q_0 = [q]_{\sim_a}$, i.e., the states that agent $a$ considers possible when the system is in $q$.
\end{definition}

\begin{definition}[Data structures]
The algorithm uses the following data structures:
\begin{itemize}
\item The model;
\item A list of \emph{information sets} for agent $a$, represented by pairs $(id,Q_{id})$ where $id\in\Nat$ is the identifier of the info set, and $Q_{id}\subseteq\States$ is an abstraction class of the $\sim_a$ relation;
\item A list of \emph{partial strategies} $PStr$ represented by the following tuples: \\
    \centerline{$(id,\strat_{id},\In_{id},\RDom_{id},\Conflicts_{id},Out_{id})$} \\
    where $id$ is the identifier of the information set on which the strategy operates, $\strat_{id}$ is the current set of choices, $\In_{id}$ the set of input states, $\RDom_{id}$ is the domain of relevance of $\strat_{id}$ from $\In_{id}$, $\Conflicts_{id}$ is the current set of conflicts, and $Out_{id}$ is the set of output states, i.e., the states by which $\strat_{id}$ can pass the control to another partial strategy.
\end{itemize}
\end{definition}

The main part of the procedure is defined by Algorithms~\ref{fig:alg-syn}, \ref{fig:alg-once} and~\ref{fig:alg}.
Algorithm~\ref{fig:alg-syn} tries to generate a perfect information strategy by employing a standard algorithm, e.g., the well-known fixpoint algorithm of~\cite{Alur02ATL}. If successful, it produces:
\begin{itemize}
\item An ordered list of epistemic indistinguishability classes, also known as \emph{information sets}, for agent $a$.
The list is generated by means of depth-first search through the transition network, starting from the initial state.
Note that the information sets are restricted to the pruning of model $M$ by strategy $\sigma$, denoted $\update{M}{\fullstrat}$ in the pseudocode.
That is, only states reachable by $\sigma$ from the initial state will be taken into account when looking at potential conflicts between $a$'s choices;

\item The ordered list of partial strategies extracted from $\sigma$, following the same ordering that was established for the information sets.
\end{itemize}
After that, Algorithm~\ref{fig:alg-syn} calls Algorithm~\ref{fig:alg}.

Algorithm~\ref{fig:alg} proceeds in cycles. In each cycle it calls Algorithm~\ref{fig:alg-once}, which optimizes the partial strategies one by one, following the ordering established by Algorithm~\ref{fig:alg-syn}. Moreover, each partial strategy is optimized first with respect to the uniformity-dominance, and then according to the outcome-dominance; this proceeds in a loop until a fixpoint is found.
Algorithm~\ref{fig:alg} terminates when no improvement has been seen in the latest iteration, or the timeout is reached.

It is worth emphasizing that, except for the first phase (generation of a perfect information strategy), this is an anytime algorithm.
It means that the procedure will return \emph{some} strategy even for models whose size is beyond grasp for optimal model checking algorithms.
This is a clear advantage over the existing approaches~\cite{Lomuscio06uniform,Lomuscio17mcmas,Busard15reasoning,Busard17phd,Pilecki17smc,Jamroga19fixpApprox-aij,Kurpiewski19domination}
where the algorithms typically provide no output even for relatively small models.

\section{Experimental Evaluation}\label{sec:experiments}
We evaluate the algorithm of Section~\ref{sec:algorithm} through experiments with two classes of models: randomly generated models and the Drones benchmark of~\cite{Kurpiewski19domination}.

\subsection{First Benchmark: Random Models}

As the first benchmark for our experiments, we use randomly generated models of a given size. The models represent a single agent playing against a nondeterministic environment. The models are generated according to the following procedure. First, we begin by generating a directed graph with several, randomly chosen, connections. The size of the graph is given by the parameter. Subsequently, we introduce additional connections between randomly selected nodes from distinct paths, in order to increase the complexity of the resulting model. Winning states are selected from the set containing the final states from each of the paths.

Once the graph is generated, it is used to construct the model. Each node represents a unique state, and a connection between two nodes indicates the presence of at least one transition between them. The transitions are generated using the following approach: for each node, a subset of outgoing connections is randomly chosen. From this subset, a set of transitions is created with actions selected randomly. As a result, some transitions will be influenced not only by the agent but also by the nondeterministic environment. This process is repeated multiple times. In the final step of the model generation algorithm, atomic propositions are randomly assigned to states, and epistemic classes are generated at random.

The number of connections, actions, winning states and epistemic classes is given as the function of the number of states in the model.

\subsection{Second Benchmark: Drone Model}

As the second benchmark we use the Drone Model from~\cite{Kurpiewski19domination} with some minor modifications.
In this scenario drones are used to measure the air quality in the specified area.
The motivation is clear, as nowadays many cities face a problem of air pollution.

A model is described using three variables:
\begin{itemize}
	\item Number of drones;
	\item Initial energy for each drone;
	\item Map size, i.e., the number of places in the area.
\end{itemize}

Every drone is equipped with a limited battery, initially charged to some energy level. Each action that the drone performs uses one energy unit. When the battery is depleted, the drone lands on the ground and must be picked up.

In our scenario, in contrast to the original one, the map is randomly generated as a directed graph. This introduces randomization into the model generation process, enabling us to thoroughly test our algorithms.
It is guaranteed that the graph is connected, and each node can be reached from the initial one. Furthermore, each node has no more than four neighbors: one for each direction of the world. Along with the map, pollution readings are also randomly generated and assigned to each place. Readings can have one of the two values: pollution or no pollution.

Each drone holds information about its current energy level, the set of already visited places and its current position on the map. When in a coalition, the drones share their data between themselves, as it is often done in real-life applications.
The indistinguishability relations are given by a faulty GPS mechanism: some of the places on the map are indistinguishable for the drone. In that way, epistemic classes are defined.

At each step, the drone can perform one of the listed actions:
\begin{itemize}
	\item Fly in one of four directions: North, West, South or East;
	\item Wait, i.e., stay in the current place.
\end{itemize}
As mentioned before, each action costs the drone one unit of its energy level.
Due to the unpredictable nature of the wind, when performing the \emph{fly} action the drone can be carried away to a different place from the one it intended.

\begin{figure*}[t]\centering
  \scalebox{0.90}{
\begin{tabular}{|c|c|c|c|c|c|c|c|c|c|c|c|c|} \hline
    &                       & \multicolumn{3}{c|}{Strategy  Perfect Info} & \multicolumn{4}{c|}{Simplified\  Strategy}          & \multicolumn{2}{c|}{Approximation}                    & \multicolumn{2}{c|}{Domino  DFS} \\ \hline
\#st & G. time & G. time           & \#str           & \#ep           & G. time                                      &  \#str & \#ep & \%ir & Time & Conclusive & Time                                 & True                              \\ \hline
10         & 0.014              & 0.031                 & 10                         & 5                            & 42.033                                           & 7                & 0    & 100\%              & 0.018   & 50\%      & 0.57                             & 100\%                               \\ \hline
100        & 0.176            & 0.546                 & 92                         & 61                           & 60.210                                            & 83               & 0  & 100\%               & 0.519  & 20\%      & 90                                  & TIMEOUT                             \\ \hline
1000       & 9.401            & 22.001                & 882                        & 629                          & 61.865 & 780              & 0    & 100\%            & 3.136   & 0\%       & 90                                   & TIMEOUT  \\ \hline
\end{tabular}
  }
\caption{Random Model results with logarithmic epistemic classes}
\label{fig:random-small}
\end{figure*}

\begin{figure*}[t]\centering
  \scalebox{0.90}{
\begin{tabular}{|c|c|c|c|c|c|c|c|c|c|c|c|c|} \hline
    &                       & \multicolumn{3}{c|}{Strategy  Perfect Info} & \multicolumn{4}{c|}{Simplified\  Strategy}          & \multicolumn{2}{c|}{Approximation}                    & \multicolumn{2}{c|}{Domino  DFS} \\ \hline
\#st & G. time & G. time           & \#str           & \#ep           & G. time                                      &  \#str & \#ep & \%ir & Time & Conclusive & Time                                 & True                              \\ \hline
10         & 0.009              & 0.023                 & 10                         & 5                            & 24.048                                           & 6                & 3     & 20\%            & 0.017   & 80\%         & 1.14                            & 100\%                               \\ \hline
100        & 0.202             & 0.489                 & 94                         & 58                           & 54.253                                            & 66               & 36      & 10\%           & 0.197   & 0\%         & 90                                   & TIMEOUT                             \\ \hline
1000       & 10.817            & 25.239                & 917                        & 584                          & 61.496 & 614              & 347           & 10\%      & 2.647   & 0\%           & 90                                   & TIMEOUT  \\ \hline
\end{tabular}
  }
\caption{Random Model results with linear epistemic classes}
\label{fig:random-big}
\end{figure*}

\begin{figure*}[t]\centering
  \scalebox{0.85}{
\begin{tabular}{|c|c|c|c|c|c|c|c|c|c|c|c|c|c|} \hline
    &        &                       & \multicolumn{3}{c|}{Strategy  Perfect Info} & \multicolumn{4}{c|}{Simplified\  Strategy}          & \multicolumn{2}{c|}{Approximation}                    & \multicolumn{2}{c|}{Domino  DFS} \\ \hline
 Map & \#st & G. time & G. time           & \#str           & \#ep           & G. time                                      &  \#str & \#ep & \%ir & Time & Conclusive & Time                                 & True                              \\ \hline
 5 & 330         & 0.078             & 0.043                 & 38                         & 13                            & 36.003                                           & 13               & 1      & 60\%            & 0.036   & 0\%        & 9.012                            & 90\%                               \\ \hline
 10 & 10648         & 3.420             & 1.284                 & 74                         & 33                            & 42.478                                           & 30               & 5     &     60\%         & 1.895   & 0\%         & 90                           & TIMEOUT                               \\ \hline
 \end{tabular} 
  }
\caption{Drone Model results}
\label{fig:drone}
\end{figure*}

\subsection{Running the Experiments}

In the experiments, we have tested 10 cases for each benchmark and each configuration, and collected the average results. Due to the randomized nature of the models, it was possible that the model generation produces a structure where no winning perfect information strategy existed. Such models were disregarded in the output of the experiments. We note in passing that, for the Randomized Model benchmark, winning perfect information strategies existed in approximately 70\% of cases.

For each test case, first the perfect information strategy was randomly chosen, and then its optimized version was generated according to Algorithm~\ref{fig:alg}. We compared our results with two other methods: fixpoint approximation from \cite{Jamroga19fixpApprox-aij} and DominoDFS from \cite{Kurpiewski19domination}. Both algorithms were implemented in Python as well as the strategy optimization algorithm. The code is available online at \url{https://github.com/blackbat13/stv}.

Random Model was tested in two different configurations that differ only by the function that binds the size of epistemic classes. In the first configuration, the maximum size of the epistemic classes was given by $\log_2 n$, where $n$ is the the number of states in the model. In the second configuration, the size of the epistemic classes was at most $10\%\ n$, i.e., linear wrt to the size of the state space.

For both benchmarks, only singleton coalitions were considered.
In particular, for Drone Model, we only generated models with a single drone acting against the environment.\footnote{
  Preliminary experiments for coalitions of drones are presented in Section~\ref{sec:coalitions}. }
The initial energy of the drone was defined as the number of places in the map times two, in order to increase the likelihood of generating a model in which the drone can visit all the places on the randomly generated map.

The experiments were conducted on an Intel Core i7-6700 CPU with dynamic clock speed of 2.60--3.50 GHz, 32 GB RAM, running under 64bit Linux Debian.

\subsection{Results}

The output of the experiments is presented in Figures~\ref{fig:random-small}, \ref{fig:random-big} and \ref{fig:drone}.
Figures~\ref{fig:random-small} and \ref{fig:random-big} present the results for the Random Model benchmark; Figure \ref{fig:drone} presents the results for the Drone Model benchmark. All running times are given in seconds. The timeout was set to 90 seconds. In case of strategy optimization, this was split into two parts: 30 seconds for the strategy generation, and 60 second for its optimization.

The first columns present information about the model configuration, its size and generation time. The next seven columns describe the output of our algorithms, i.e., the randomly generated strategy with perfect information and its optimized version. The last part of the tables contains the reference results from the algorithms used for comparison: lower and upper fixpoint approximation and DominoDFS method.

The table headers should be interpreted as follows:
\begin{itemize}
	\item $Map$: number of places on the map (for Drone Model);
	\item $\#st$: number of states in the model;
	\item $G. time$: generation time for the model/strategy;
	\item $\#str$: number of states reachable in the strategy;
	\item $\#ep$: number of states in which the strategy uniformity was broken;
	\item $\%ir$: percentage of cases in which optimized strategy was a uniform strategy;
	\item $Time$: time used by the Approximation/Domino DFS algorithm;
	\item \emph{Conclusive}: percentage of cases in which the result of fixpoint approximation was conclusive, i.e. when both the upper bound and the lower bound computations yield the same outcome;
	\item $True$: percentage of cases in which Domino DFS returned a winning strategy (timeout was reached in all the other cases).
\end{itemize}

As the results show, our method performed very well
in comparison to the reference algorithms.
The DominoDFS method ended mostly with timeout for larger models, and the fixpoint approximations gave mostly inconclusive results.
In contrast, our optimized strategies obtained pretty good elimination of conflicts, and in many cases produced ideal, i.e., fully uniform strategies.

The results also show clearly that our optimization algorithm works best in situations when the size of the epistemic classes is relatively small. For the logarithmic size of the epistemic classes, the optimized strategy was always a uniform strategy (!). As for the setting with the linear size, the optimization-based algorithm was not as good, but still gave a reduction of conflicts of about 40\%.
Even in that case, it produced ideal strategies in $10-20\%$ of instances.
It is also worth pointing out that, for the Drone benchmark, our optimization returned a uniform strategy in about 60\% cases.

We note, again, that our algorithm is an anytime algorithm, which means that it always returns \emph{some} strategy, regardless of the given timeout.

\section{Coalitional Strategies}\label{sec:coalitions}
\begin{algorithm}[t]
    \caption{Optimization algorithm for coalition $optimize\_coal(PStr, A)$}
    \begin{algorithmic}
    \REPEAT
        \STATE{$OldPStr := PStr$}
        \FOR {$agent$ in $A$}
            \STATE{$Pstr := optimize\_once(PStr)$}
        \ENDFOR
    \UNTIL{timeout or (PStr = OldPStr)}
    \RETURN{$PStr$}
    \end{algorithmic}
    \label{fig:alg-coal}
\end{algorithm}

So far, we have focused on the synthesis of individual strategies.
In fact, our synthesis algorithm in Section~\ref{sec:algorithm} works only for singleton coalitions.
This is because it relies on the fact that the domains of partial strategies are closed with respect to indistinguishability relations of the involved agents.
While such a closure is guaranteed for information sets of single agent, the {union} of information sets of several agents typically does not satisfy the property.

One way out is to \emph{define} the domains of partial strategies by the closure.
The domains would in that case correspond to common knowledge neighborhoods for the coalition.
Unfortunately, this will not work well in practice: for most models, the common knowledge closure will produce the whole state space, and thus make the computation infeasible.

Another simple idea is to optimize coalitional strategies agent-wise, alternating between the agents.
In that case, we optimize the individual strategies being parts of $\sigma_A$ one by one, using the optimization template from Section~\ref{sec:algorithm}. The resulting procedure is presented as Algorithm~\ref{fig:alg-coal}.

The output of our experimental evaluation for synthesis of coalitional strategies is presented in Figure~\ref{fig:drone-coal}.
For the experiments, the Drone benchmark was selected with coalition of two drone agents.
As the results show, our algorithm obtained a high level of optimization of the initial, perfect information, strategy.
Most importantly, the procedure produced ideal strategies in $60\%$ and $40\%$ of the instances, respectively,
thus providing a conclusive answer to the model checking question in about half of the cases.

\begin{figure}[t]\centering
    \scalebox{0.95}{
\begin{tabular}{|c|c|c|c|c|c|c|c|c|c|} \hline
    &        &                       & \multicolumn{3}{c|}{Strategy  Perfect Info} & \multicolumn{4}{c|}{Simplified\  Strategy}           \\ \hline
 Map & \#st & G. time & G. time           & \#st           & \#ep           & G. time                                      &  \#st & \#ep & \%ir                              \\ \hline
 3 & 667         & 0.85             & 0.397                 & 35                         & 11                            & 12.031                                           & 9               & 1      & 60\%                  \\ \hline        
 5 & 31122         & 69.265             & 107.428                 & 587                         & 728                            & 60.8                                           & 87               & 58     &     40\%               \\ \hline    
 \end{tabular} 
    }
\caption{Drone model results for coalitions}
\label{fig:drone-coal}
\end{figure}

\section{Conclusions}\label{sec:conclusions}

In this paper, we propose an anytime algorithm to synthesize ``reasonably good'' strategies for reachability goals under imperfect information.
The idea is to first generate a surely winning strategy with \emph{perfect information}, and then iteratively improve it with respect to its uniformity level and the tightness of its outcome set.
We evaluate the algorithm experimentally on two classes of models: randomly generated ones and ones modeling a group of drones patrolling for air pollution.
The results show high optimization rates, especially for models with relatively small indistinguishability classes.
For such models, the procedure produced ideal strategies in a large fraction of the instances,
thus providing a conclusive answer to the model checking question.

The fact that our method works well for models with small epistemic classes suggests that it should complement, rather than compete, with methods based on search through the space of uniform strategies (which usually work better for models with \emph{large} information sets).
Depending on the kind of the model, a suitable algorithm should be used.

\subsubsection{Acknowledgements}
The work was supported by NCBR Poland and FNR Luxembourg under the PolLux/FNR-CORE projects STV (POLLUX-VII/1/2019 and C18/IS/12685695/IS/STV/Ryan), SpaceVote (POLLUX-XI/14/SpaceVote/2023 and C22/IS/17232062/SpaceVote) and PABLO (C21/IS/16326754/PABLO). The work of Damian Kurpiewski was also supported by the CNRS IEA project MoSART.

\clearpage
\bibliographystyle{splncs04}

\begin{thebibliography}{10}
    \providecommand{\url}[1]{\texttt{#1}}
    \providecommand{\urlprefix}{URL }
    \providecommand{\doi}[1]{https://doi.org/#1}
    
    \bibitem{Alur00mocha}
    Alur, R., de~Alfaro, L., Henzinger, T.A., Krishnan, S., Mang, F., Qadeer, S.,
      Rajamani, S., Tasiran, S.: {MOCHA}: Modularity in model checking. Tech. rep.,
      University of Berkeley (2000)
    
    \bibitem{Alur97ATL}
    Alur, R., Henzinger, T.A., Kupferman, O.: {A}lternating-time {T}emporal
      {L}ogic. In: Proceedings of the 38th Annual Symposium on Foundations of
      Computer Science (FOCS). pp. 100--109. IEEE Computer Society Press (1997)
    
    \bibitem{Alur02ATL}
    Alur, R., Henzinger, T.A., Kupferman, O.: {A}lternating-time {T}emporal
      {L}ogic. Journal of the ACM  \textbf{49},  672--713 (2002).
      \doi{10.1145/585265.585270}
    
    \bibitem{Aminof18gradedSL}
    Aminof, B., Malvone, V., Murano, A., Rubin, S.: Graded modalities in {Strategy
      Logic}. Information and Computation  \textbf{261},  634--649 (2018).
      \doi{10.1016/j.ic.2018.02.022}
    
    \bibitem{Jamroga17bisimATLir}
    Belardinelli, F., Condurache, R., Dima, C., Jamroga, W., Jones, A.:
      Bisimulations for verification of strategic abilities with application to
      {ThreeBallot} voting protocol. In: Proceedings of the 16th International
      Conference on Autonomous Agents and Multiagent Systems (AAMAS). pp.
      1286--1295. IFAAMAS (2017)
    
    \bibitem{Belardinelli17abstraction}
    Belardinelli, F., Lomuscio, A.: Agent-based abstractions for verifying
      alternating-time temporal logic with imperfect information. In: Proceedings
      of {AAMAS}. pp. 1259--1267. {ACM} (2017)
    
    \bibitem{Berthon17sl-imperfectinfo}
    Berthon, R., Maubert, B., Murano, A., Rubin, S., Vardi, M.Y.: Strategy logic
      with imperfect information. In: Proceedings of {LICS}. pp. 1--12 (2017).
      \doi{10.1109/LICS.2017.8005136}
    
    \bibitem{BerthonMMRV21}
    Berthon, R., Maubert, B., Murano, A., Rubin, S., Vardi, M.Y.: Strategy logic
      with imperfect information. {ACM} Trans. Comput. Log.  \textbf{22}(1),
      5:1--5:51 (2021). \doi{10.1145/3427955},
      \url{https://doi.org/10.1145/3427955}
    
    \bibitem{Bryant86boolean}
    Bryant, R.E.: Graph-based algorithms for boolean function manipulation. IEEE
      Trans. on Computers  \textbf{35}(8),  677--691 (1986)
    
    \bibitem{Bulling10verification}
    Bulling, N., Dix, J., Jamroga, W.: Model checking logics of strategic ability:
      Complexity. In: Dastani, M., Hindriks, K., Meyer, J.J. (eds.) Specification
      and Verification of Multi-Agent Systems, pp. 125--159. Springer (2010)
    
    \bibitem{Bulling09patl-fundamenta}
    Bulling, N., Jamroga, W.: What agents can probably enforce. Fundamenta
      Informaticae  \textbf{93}(1-3),  81--96 (2009)
    
    \bibitem{Bulling11mu-ijcai}
    Bulling, N., Jamroga, W.: Alternating epistemic mu-calculus. In: Proceedings of
      {IJCAI-11}. pp. 109--114 (2011)
    
    \bibitem{Burch90symbolic-mcheck}
    Burch, J.R., Clarke, E.M., McMillan, K.L., Dill, D.L., Hwang, L.J.: Symbolic
      model checking: 10{\^{}}20 states and beyond. In: Proc. of 4th Ann. {IEEE}
      Symp. on Logic in Computer Science ({LICS}). pp. 428--439. {IEEE} Computer
      Society (1990)
    
    \bibitem{Busard17phd}
    Busard, S.: Symbolic Model Checking of Multi-Modal Logics: Uniform Strategies
      and Rich Explanations. Ph.D. thesis, Universite Catholique de Louvain (2017)
    
    \bibitem{Busard14improving}
    Busard, S., Pecheur, C., Qu, H., Raimondi, F.: Improving the model checking of
      strategies under partial observability and fairness constraints. In: Formal
      Methods and Software Engineering, Lecture Notes in Computer Science,
      vol.~8829, pp. 27--42. Springer (2014). \doi{10.1007/978-3-319-11737-9_3}
    
    \bibitem{Busard15reasoning}
    Busard, S., Pecheur, C., Qu, H., Raimondi, F.: Reasoning about memoryless
      strategies under partial observability and unconditional fairness
      constraints. Information and Computation  \textbf{242},  128--156 (2015).
      \doi{10.1016/j.ic.2015.03.014}
    
    \bibitem{Calta10finding}
    Calta, J., Shkatov, D., Schlingloff, B.H.: Finding uniform strategies for
      multi-agent systems. In: Proceedings of Computational Logic in Multi-Agent
      Systems (CLIMA). Lecture Notes in Computer Science, vol.~6245, pp. 135--152.
      Springer (2010)
    
    \bibitem{Cermak14mcheckSL}
    Cermak, P., Lomuscio, A., Mogavero, F., Murano, A.: {MCMAS-SLK}: A model
      checker for the verification of strategy logic specifications. In: Proc. of
      Computer Aided Verification (CAV). Lecture Notes in Computer Science,
      vol.~8559, pp. 525--532. Springer (2014)
    
    \bibitem{Cermak15mcmas-sl-one-goal}
    Cerm{\'{a}}k, P., Lomuscio, A., Murano, A.: Verifying and synthesising
      multi-agent systems against one-goal strategy logic specifications. In:
      Proceedings of {AAAI}. pp. 2038--2044 (2015)
    
    \bibitem{Chatterjee07omegareg}
    Chatterjee, K., Doyen, L., Henzinger, T., Raskin, J.F.: Algorithms for
      omega-regular games of incomplete information. Logical Methods in Computer
      Science  \textbf{3}(3) (2007)
    
    \bibitem{Clarke81ctl}
    Clarke, E., Emerson, E.: Design and synthesis of synchronization skeletons
      using branching time temporal logic. In: Proceedings of Logics of Programs
      Workshop. Lecture Notes in Computer Science, vol.~131, pp. 52--71 (1981)
    
    \bibitem{Dembinski03verics}
    Dembi\'nski, P., Janowska, A., Janowski, P., Penczek, W., P\'o{\l}rola, A.,
      Szreter, M., Wo\'zna, B., Zbrzezny, A.: Verics: A tool for verifying timed
      automata and estelle specifications. In: Proceedings of the of the 9th Int.
      Conf. on Tools and Algorithms for Construction and Analysis of Systems
      (TACAS'03), Lecture Notes in Computer Science, vol.~2619, pp. 278--283.
      Springer (2003)
    
    \bibitem{Dima14mucalc}
    Dima, C., Maubert, B., Pinchinat, S.: The expressive power of epistemic
      $\mu$-calculus. CoRR  \textbf{abs/1407.5166} (2014)
    
    \bibitem{Dima15fallmu}
    Dima, C., Maubert, B., Pinchinat, S.: Relating paths in transition systems: The
      fall of the modal mu-calculus. In: Proceedings of Mathematical Foundations of
      Computer Science (MFCS). Lecture Notes in Computer Science, vol.~9234, pp.
      179--191. Springer (2015). \doi{10.1007/978-3-662-48057-1_14}
    
    \bibitem{Dima11undecidable}
    Dima, C., Tiplea, F.: Model-checking {ATL} under imperfect information and
      perfect recall semantics is undecidable. CoRR  \textbf{abs/1102.4225} (2011)
    
    \bibitem{Doyen11games}
    Doyen, L., Raskin, J.F.: Games with imperfect information: Theory and
      algorithms. In: Lecture Notes in Game Theory for Computer Scientists, pp.
      185--212. Cambridge University Press (2011)
    
    \bibitem{Gammie04mck}
    Gammie, P., Meyden, R.: {MCK}: Model checking the logic of knowledge. In: Proc.
      of the 16th Int. Conf. on Computer Aided Verification (CAV'04). LNCS,
      vol.~3114, pp. 479--483. Springer-Verlag (2004)
    
    \bibitem{Huang14symbolic-epist}
    Huang, X., van~der Meyden, R.: Symbolic model checking epistemic strategy
      logic. In: Proceedings of AAAI Conference on Artificial Intelligence. pp.
      1426--1432 (2014)
    
    \bibitem{Jamroga15specificationMAS}
    Jamroga, W.: Logical Methods for Specification and Verification of Multi-Agent
      Systems. ICS PAS Publishing House (2015)
    
    \bibitem{Jamroga06atlir-eumas}
    Jamroga, W., Dix, J.: Model checking {ATL$_{ir}$} is indeed
      {$\Delta_2^P$}-complete. In: Proceedings of EUMAS. {CEUR} Workshop
      Proceedings, vol.~223 (2006)
    
    \bibitem{Jamroga18por}
    Jamroga, W., Penczek, W., Dembi\'nski, P., Mazurkiewicz, A.: Towards partial
      order reductions for strategic ability. In: Proceedings of the 17th
      International Conference on Autonomous Agents and Multiagent Systems (AAMAS).
      pp. 156--165. IFAAMAS (2018)
    
    \bibitem{Jamroga19somethings}
    Jamroga, W., Knapik, M.: Some things are easier for the dumb and the bright
      ones (beware the average!). In: Proceedings of the Twenty-Eighth
      International Joint Conference on Artificial Intelligence {IJCAI}. pp.
      1734--1740 (2019). \doi{10.24963/ijcai.2019/240}
    
    \bibitem{Jamroga19fixpApprox-aij}
    Jamroga, W., Knapik, M., Kurpiewski, D., Mikulski, {\L}.: Approximate
      verification of strategic abilities under imperfect information. Artificial
      Intelligence  \textbf{277} (2019). \doi{10.1016/j.artint.2019.103172}
    
    \bibitem{Kurpiewski19stv-demo}
    Kurpiewski, D., Jamroga, W., Knapik, M.: {STV}: Model checking for strategies
      under imperfect information. In: Proceedings of the 18th International
      Conference on Autonomous Agents and Multiagent Systems AAMAS 2019. pp.
      2372--2374. IFAAMAS (2019)
    
    \bibitem{Kurpiewski19domination}
    Kurpiewski, D., Knapik, M., Jamroga, W.: On domination and control in strategic
      ability. In: Proceedings of the 18th International Conference on Autonomous
      Agents and Multiagent Systems AAMAS 2019. pp. 197--205. IFAAMAS (2019)
    
    \bibitem{Kurpiewski21stv-demo}
    Kurpiewski, D., Pazderski, W., Jamroga, W., Kim, Y.: {STV+Reductions}: Towards
      practical verification of strategic ability using model reductions. In:
      Proceedings of AAMAS. pp. 1770--1772. ACM (2021)
    
    \bibitem{Lomuscio17mcmas}
    Lomuscio, A., Qu, H., Raimondi, F.: {MCMAS}: An open-source model checker for
      the verification of multi-agent systems. International Journal on Software
      Tools for Technology Transfer  \textbf{19}(1),  9--30 (2017).
      \doi{10.1007/s10009-015-0378-x}
    
    \bibitem{Lomuscio06uniform}
    Lomuscio, A., Raimondi, F.: Model checking knowledge, strategies, and games in
      multi-agent systems. In: Proceedings of International Joint Conference on
      Autonomous Agents and Multiagent Systems (AAMAS). pp. 161--168 (2006).
      \doi{10.1145/1160633.1160660}
    
    \bibitem{Mogavero14behavioral}
    Mogavero, F., Murano, A., Perelli, G., Vardi, M.: Reasoning about strategies:
      On the model-checking problem. ACM Transactions on Computational Logic
      \textbf{15}(4),  1--42 (2014)
    
    \bibitem{Mogavero10stratLogic}
    Mogavero, F., Murano, A., Vardi, M.: Reasoning about strategies. In:
      Proceedings of FSTTCS. pp. 133--144 (2010)
    
    \bibitem{Peterson79alternation}
    Peterson, G., Reif, J.: Multiple-person alternation. In: Proceedings of the
      20th Annual Symposium on Foundations of Computer Science (FOCS). pp.
      348--363. IEEE Computer Society Press (1979)
    
    \bibitem{Pilecki14synthesis}
    Pilecki, J., Bednarczyk, M., Jamroga, W.: Synthesis and verification of uniform
      strategies for multi-agent systems. In: Proceedings of {CLIMA XV}. Lecture
      Notes in Computer Science, vol.~8624, pp. 166--182. Springer (2014)
    
    \bibitem{Pilecki17smc}
    Pilecki, J., Bednarczyk, M., Jamroga, W.: {SMC}: Synthesis of uniform
      strategies and verification of strategic ability for multi-agent systems.
      Journal of Logic and Computation  \textbf{27}(7),  1871--1895 (2017).
      \doi{10.1093/logcom/exw032}
    
    \bibitem{Raimondi07obdds}
    Raimondi, F., Lomuscio, A.: Automatic verification of multi-agent systems by
      model checking via ordered binary decision diagrams. J. Applied Logic
      \textbf{5}(2),  235--251 (2007)
    
    \bibitem{Schobbens04ATL}
    Schobbens, P.Y.: Alternating-time logic with imperfect recall. Electronic Notes
      in Theoretical Computer Science  \textbf{85}(2),  82--93 (2004)
    
    \end{thebibliography}

\end{document}